\documentclass[prl,twocolumn,a4paper,superscriptaddress,showpacs]{revtex4-1}
\usepackage[utf8]{inputenc}
\usepackage[T1]{fontenc}
\usepackage[english]{babel}
\usepackage{graphicx,amsmath,bm,natbib,dsfont,color,ulem}

\def\bra#1{\mathinner{\langle{#1}|}}
\def\ket#1{\mathinner{|{#1}\rangle}}
\def\mn#1{\langle #1 \rangle}
\def\prjct#1{\mathinner{|{#1}\rangle}\!\!\mathinner{\langle{#1}|}}
\newcommand{\coh}[2]{\mathinner{|{#1}\rangle}\!\!\mathinner{\langle{#2}|}}

\def\text#1{\textrm{#1}}

\def\eq{\begin{equation}}
\def\eeq{\end{equation}}

\begin{document}

\title{Two-mode squeezed states as Schr\"odinger-cat-like states}
\author{E.~Oudot}
\affiliation{Department of Physics, University of Basel, CH-4056 Basel, Switzerland}
\affiliation{Group of Applied Physics, University of Geneva, CH-1211 Geneva 4, Switzerland}
\author{P.~Sekatski}
\affiliation{Institut f\"ur Theoretische Physik, Universit\"at Innsbruck, Technikerstr. 25, A-6020 Innsbruck, Austria}
\author{F.~Fr\"owis}
\affiliation{Group of Applied Physics, University of Geneva, CH-1211 Geneva 4, Switzerland}
\author{N.~Gisin}
\affiliation{Group of Applied Physics, University of Geneva, CH-1211 Geneva 4, Switzerland}
\author{N.~Sangouard}
\affiliation{Department of Physics, University of Basel, CH-4056 Basel, Switzerland}

\begin{abstract}

In recent years, there has been an increased interest in the generation of superposition of coherent states with opposite phases, the so-called photonic Schr\"odinger-cat states. These experiments are very challenging and so far, cats involving small photon numbers only have been implemented. Here, we propose to consider two-mode squeezed states as examples of a Schr\"odinger-cat-like state. In particular, we are interested in several criteria aiming to identify quantum states that are macroscopic superpositions in a more general sense. We show how these criteria can be extended to continuous variable entangled states. We apply them to various squeezed states, argue that two-mode squeezed vacuum states belong to a class of general Schr\"odinger-cat states and compare the size of states obtained in several experiments. Our results not only promote two-mode squeezed states for exploring quantum effects at the macroscopic level but also provide direct measures to evaluate their usefulness for quantum metrology.
\end{abstract}
\date{\today}
\maketitle

\paragraph{Introduction ---}
The question of what is a macroscopic quantum state has received quite a lot of attention over the last decade \cite{Frowis_Measures_2012, Jeong_Characterizations_2014a, Farrow_Classification_2014a, Frowis_Linking_2015}. The motivation is not to address a new question -- not at all, as it dates back from the early days of quantum theory \cite{Schrodinger_gegenwartige_1935} -- but rather comes from the  experimental progress, now allowing one to harness large systems while highlighting their quantum nature. Quantum optics experiments reporting on squeezing operations provide a nice example. They are obtained from a $\chi^2$-nonlinearity and can result in largely entangled states. The entanglement can further be detected with homodyne detections, by means of the Duan -- Simon criterion  \cite{Duan_Inseparability_2000, Simon_Peres-Horodecki_2000}. When the $\chi^2$-nonlinearity is seeded by coherent states and/or embedded in a high finesse cavity, entanglement in squeezed states can be demonstrated with a huge number of photons -- so huge that they can be detected with classical power-meters \cite{Zhang_Experimental_2000, Silberhorn_Generation_2001, Bowen_Experimental_2003, Villar_Generation_2005, Keller_Experimental_2008, Wang_Experimental_2010, Eberle_Experimental_2013}. This naturally raises the question of whether squeezed states have macroscopic quantum features -- a question of deep relevance because so far squeezed states have been combined with conditional detections \cite{Ourjoumtsev_Generating_2006, Neergaard-Nielsen_Generation_2006, Wakui_Photon_2007, Bruno_displacement_2013, Lvovsky_Observation_2013, Morin_Remote_2014, Jeong_Generation_2014} for exploring quantum effects in many photon states.

In the literature, there exist different criteria for quantifying the macroscopic quantumness \cite{Dur_effective_2002, Shimizu_stability_2002, Bjork_size_2004, Shimizu_Detection_2005,Cavalcanti_Signatures_2006, Korsbakken_Measurement-based_2007, Marquardt_Measuring_2008, Lee_Quantification_2011, Frowis_Measures_2012, Nimmrichter_Macroscopicity_2013, Sekatski_Size_2014, Laghaout_Assessments_2014a}. Typically, this includes a definition that assigns to a quantum state a number, which is here called \textit{effective size} (or simply \textit{size}). Surprisingly, none of them unambiguously applies to two-mode squeezed states and, at the same time, is able to compare their size to those of other states. These criteria can be grouped into two categories. The first one addresses the question of whether a two component superposition $\ket{\phi_0}+\ket{\phi_1}$ is macroscopic, i.e., whether  $|\phi_0\rangle$ and $|\phi_1\rangle$ are \textit{macroscopically distinct}. For example, the proposal of Ref.~\cite{Korsbakken_Measurement-based_2007} states that two spin states are macroscopically distinct if they can be distinguished from a small number of their spins -- as a dead cat and an alive cat can be distinguished from a small number of their cells. We can also refer to the proposals of Ref.~\cite{Sekatski_Proposal_2012, Sekatski_Size_2014} defining two states as being macroscopically distinct if they can be distinguished with a coarse-grained measurement -- as a dead cat and an alive cat can be distinguished with a detector having a very limited resolution. The second category aims to identify quantum states that are able to show some kind of \textit{macroscopic quantum effect}. This term characterizes experimental evidence that can not be explained by an accumulated quantum effect originated at the microscopic level of the system. For pure states, a large variance with respect to given observables and Hamiltonians is a sufficient signature for quantum fluctuations that are persistent on a macroscopic level. For mixed states, one typically uses a convex function that reduces to the variance for pure states. For example, the proposal of Ref.~\cite{Frowis_Measures_2012} shows how the notion of macroscopicity can be linked to the so-called quantum Fisher information \cite{Braunstein_Statistical_1994}. Focusing on photonic states, both groups have strong limitations. The first ones only apply to states of the form $\ket{\phi_0} + \ket{\phi_1}$ and cannot be used directly to measure the size of continuous variable (cv) states. The second category does not focus on a specific state structure but an unambiguous extension to multimode states is missing \footnote{Note that the recent work presented in \cite{Laghaout_Assessments_2014a} in an extension of the measures of the first categories to non-qubit states but it is unclear how to extend it properly to the multimode case}.

In this letter, we show how representative measures of each groups can be extended to cv entangled states. These extensions allow one to characterize the macroscopicness of two-mode squeezed states. In particular, we prove that the effective size of two-mode squeezed vacuum states (with $N$ mean photons) is basically the same as superpositions of coherent states with opposite phases $|\alpha\rangle + |-\alpha\rangle$ and $|\alpha|^2 \tanh |\alpha|^2 = N$; but with the great advantage that they are much easier to create. The tools that we propose allow one to bound the size of states obtained experimentally as well as their usefulness for parameter estimations beyond the classical limit. Aside from their fundamental interest, our results thus have important applications for quantum metrology. \\

\paragraph{Two-mode vacuum squeezed states ---} 
As an example of two-mode squeezed states, let us consider the two-mode squeezed vacuum. It is obtained from a parametric process in which photons from a pump laser decay spontaneously into photon pairs -- one in mode 1,  its twin in mode 2 -- while preserving energy and momentum. The corresponding propagator $\bar S(g)=e^{g (a_1 a_2-a^\dag_1 a^\dag_2)}$ with squeezing parameter $g,$ applies straightforwardly on the vacuum if written in the normal order. This results in
\begin{equation}
\label{2modesqueezed}
\psi_{\text{tms}}=(1-\tanh^2 g)^{\frac{1}{2}} \-\ e^{\tanh g \-\ a^\dag_1 a^\dag_2} |00\rangle.
\end{equation}  
The mean photon number in both mode is $N=2\text{tr}(a^\dag_1 a_1 |\psi_{\text{tms}}\rangle\langle \psi_{\text{tms}}|)=2\sinh^2 g.$ Furthermore, the variance of the observable $\bar X_1^\varphi - \bar X_2^\phi$ where $\bar X_i^\theta = \frac{1}{\sqrt{2}}\left(a_i e^{i\theta} + a_i^\dag e^{-i\theta}\right)$ is given by 
\begin{equation}
V_{\psi_{\text{tms}}} (\bar X_1^\varphi - \bar X_2^\phi)=\cosh 2g - \sinh 2g \-\ \cos(\varphi+\phi).
\end{equation}
This indicates that the quadratures $\bar X_1^0$ -- $\bar X_2^0$ are correlated whereas $\bar X_1^{\pi/2}$ -- $\bar X_2^{\pi/2}$ are anti-correlated. The quantum nature of these correlations can be revealed through the Duan -- Simon criterion \cite{Duan_Inseparability_2000, Simon_Peres-Horodecki_2000} which states that for any bipartite separable states and any real parameter $a$
\begin{eqnarray}
\nonumber
&& V_{sep}\left (|a| \bar X_1^{\phi} + \frac{1}{a} \bar X_2^{\Phi}\right) + V_{sep} \left(|a| \bar X_1^{\phi'} - \frac{1}{a} \bar X_2^{\Phi'} \right) \\
\nonumber
&& > a^2 \langle [\bar X_1^{\phi} , \bar X_1^{\phi'}]\rangle +\frac{1}{a^2}  \langle [\bar X_2^{\Phi} , \bar X_2^{\Phi'}]\rangle \\
\label{Duan-Simon}
&&\geq 2 \-\ \text{for} \-\  \phi - \phi'=\Phi - \Phi'=\frac{\pi}{2} 
\end{eqnarray}
while for a two-mode squeezed state 
$$
V_{\psi_{\text{tms}}}\left (\bar X_1^{0} - \bar X_2^{0}\right) +V_{\psi_{\text{tms}}} \left(\bar X_1^{\pi/2} + \bar X_2^{\pi/2} \right) = 2 e^{-2g}.
$$
The questions that are at the core of this letter are: How to evaluate the size of this kind of states? Are their effective size comparable to other photonic states?\\

\paragraph{Macroscopic distinctness for cv states ---} While several definitions have been proposed to identify states that are macroscopically distinct \cite{Bjork_size_2004, Shimizu_Detection_2005, Korsbakken_Measurement-based_2007, Marquardt_Measuring_2008, Sekatski_Size_2014}, we here focus on the proposal of Ref.~\cite{Sekatski_Size_2014} based on coarse-grained measurements. This choice is arbitrary to some extent. Note, however, that the extension that we propose below easily applies to the measure of Ref.~\cite{Korsbakken_Measurement-based_2007}. The extension of measures of Refs. \cite{Marquardt_Measuring_2008, Shimizu_Detection_2005} to two-mode squeezed states is less obvious as they primarily address spin systems but the link between measures for spins and photons presented in \cite{Frowis_Linking_2015} might be the way to proceed. 

The basic principle of the measure of macroscopicity based on coarse-grained measurement is simple. It can be seen as a game where Alice chooses a state in the set $\{|\phi_0\rangle, |\phi_1\rangle\}$ with equal a priori probabilities and sends it to Bob. Bob has to guess which one has been sent using a coarse-grained measurement only. It can be any measurement provided that its resolution is limited. The quantum superposition state $|\phi_0\rangle + |\phi_1\rangle$ is qualified macroscopic if Bob wins the game with a detector having no microscopic resolution. Concretely, if one focuses on a noisy photon counting detector for example, the size of $|\phi_0\rangle + |\phi_1\rangle$ is characterized by the noise that one can tolerate to distinguish $|\phi_0\rangle$ and $|\phi_1\rangle.$

To extend this measure to cv states, we can mimic its original idea by introducing a 50/50 binning of measurement outcomes. For a two-mode squeezed vacuum state in particular, Alice measures her mode with a given quadrature and bins the result with respect to its sign. As Alice's measurement is assumed to be very accurate, this binning corresponds to equiprobable projections onto two orthogonal subspaces of the measured state. Bob has to guess whether she got a positive or negative outcome by measuring his mode with a noisy measurement. The distinguishability of components that Bob receives is again given by the noise that can be tolerated to win the game. Note that the measurement of correlated quadratures maximizes the probability to correctly guess Alice's outcome. Concretely, the probability that Alice gets the result $x_1$ and Bob $x_2$ knowing that they measure the quadratures $\bar X_1^{0}$ and $\bar X_2^{0}$ is given by $|p(x_1,x_2,{\sigma})|^2=\text{tr}(|\psi_{\text{tms}}\rangle\langle \psi_{\text{tms}}| \delta (\bar X_1^{0}-x_1)  g_{\sigma}(\bar X_2^{0}-x_2))$ where $g_{\sigma} $ stands for the noise of Bob's measurement device. We assume that $g_{\sigma}$ is a Gaussian with spread $\sigma$ and zero mean. Hence, the probability that Bob correctly guesses the sign of Alice's result is given by $P_{\sigma}^{\mathrm{guess}}=\int_{0}^{+\infty} |p(x_1,x_2,{\sigma})|^2\mbox{d}x_1\mbox{d}x_2+\int_{-\infty}^{0} |p(x_1,x_2,{\sigma})|^2 \mbox{d}x_1\mbox{d}x_2.$ We find  
\begin{equation}
P_{\sigma}^{\mathrm{guess}}=\frac{1}{2} + \frac{1}{\pi} \arctan(\frac{ \sinh 2g }{ \sqrt{1+2\sigma^2 \cosh 2g }}).\end{equation}
We can access the maximum noise $\sigma_{\max}$ that Bob can tolerate to win the game with a fixed probability $P_{\sigma}^{\mathrm{guess}}$ by inverting the previous formula:
 \begin{equation}
\sigma_{\max}=\sqrt{\frac{-1+N(\frac{1}{2}+N)\mbox{cotan}^2(\frac{1}{2}-P_{\sigma}^{\mathrm{guess}})}{2+2N}}, 
\end{equation}
For comparison, the noise that can be tolerated to win a similar game with the optical Schr\"odinger-cat state $\left(|\uparrow\rangle|\alpha\rangle - |\downarrow\rangle|-\alpha\rangle \right)$ is given by \\
$$
\sigma_{\max}=\sqrt{\frac{|\alpha|^2}{\left(\text{erf}^{-1}\left(P_{\sigma}^{\mathrm{guess}}\right)\right)^2}-\frac{1}{2}}.
$$
In both cases, the noise scales like the square root of the photon number. Two-mode squeezed states and Schr\"odinger-cat states thus belong to the same class of macroscopic states.

Let us now focus on practical considerations. The observation that Alice's and Bob's $x$-quadratures of the two-mode squeezed vacuum state are ``macroscopically'' correlated (correlated at a large scale, larger then the detector's resolution) is at the heart of our generalization of the coarse-grained measure. These correlations can be revealed by measuring the joint probability distribution $|p(x_1, x_2, 0)|^2$ with accurate quadrature measurements. (For simplicity, we introduce $p(x_1, x_2)=p(x_1, x_2, 0)$ which stands for the probability amplitudes without noise.) Although this approach is sufficient to measure the size of a given state in theory, one also has to ensure that those correlations are truly quantum in practice. In mathematical terms, we can always write the state that is shared by Alice and Bob in the $x$-basis
\begin{align}
\label{state in x}
\rho = \int  p(x_1,x_2) p^*(\bar x_1,\bar x_2) f(x_1, \bar x_1, x_2, \bar x_2) \cdot\nonumber \\
 \coh{x_1,x_2}{\bar x_1, \bar x_2} d x_{1} d x_2 d\bar x_1 d\bar x_2,
\end{align}
with $\int |p(x_1,x_2)|^2 dx_1dx_2=1$ and $f(x, x, x', x') = 1$ $\forall x, x'.$ If the shared state is pure, we have $f(x_1,\bar x_1, x_2,\bar x_2) = 1$ $\forall x_1, \bar x_1, x_2, \bar x_2$ and the correlations revealed through the probability distribution $|p(x_1, x_2)|^2$ are fully quantum. The violation of the Duan-Simon criterion is then sufficient to attest the quantum nature of the state for which the size is evaluated through $\sigma_{\max}.$ But how to certify in practice that the function $f(x_1, \bar x_1, x_2, \bar x_2)$ is close to one, at least in a certain range?\\

To do so, we consider the effect of imperfect coherences (decoherence) $f(x_1, \bar x_1,x_2, \bar x_2) \neq 1$ on the observed violation of the Duan-Simon witness. Note first that the variance $V(\bar X^{0}_1 - \bar X^{0}_2)$ can be directly obtained from $|p(x_1, x_2)|^2$. For the second term required in Eq. (\ref{Duan-Simon}), we can show that the variance in presence of decoherence (see Appendix A)
\begin{eqnarray}
\nonumber
& V(\bar X^{\frac{\pi}{2}}_1 + \bar X^{\frac{\pi}{2}}_2)  = & V(\bar X^{\frac{\pi}{2}}_1 + \bar X^{\frac{\pi}{2}}_2)|_{\text{ideal}}  \\
\nonumber
&&  -\mn{ (\partial_{x_1-\bar x_1}+\partial_{x_2-\bar x_2})^2 f}.
\end{eqnarray}
equals the ideal-case variance  $V(\bar X^{\frac{\pi}{2}}_1 + \bar X^{\frac{\pi}{2}}_2)|_{\text{ideal}}$ plus a factor containing the crossed and second derivatives of $f$ 
$\mn{(\partial_{x_1-\bar x_1} f(x_1, \bar x_1,x_2, \bar x_2)}=\int dx_1 dx_2 |p(x_1,x_2)|^2 (\partial_{x_1-\bar x_1} f(x_1, \bar x_1,x_2, x_2) )|_{x_1 = \bar x_1}$, 
etc. Since $V(\bar X^{\frac{\pi}{2}}_1 + \bar X^{\frac{\pi}{2}}_2)|_{\text{ideal}}$ is positive, we get the following upper-bound on the observed variance
\eq\label{bound on f}
 - \mn{ (\partial_{x_1-\bar x_1}+\partial_{x_2-\bar x_2})^2 f} \leq V(\bar X^{\frac{\pi}{2}}_1 + \bar X^{\frac{\pi}{2}}_2)
\eeq
Note that without further assumptions, we cannot bound the range $\delta$ for which $f(x_1,x_1+\delta,x_2,x_2+\delta)$ stays close to one. In words, even if the state of Alice and Bob largely violate the Duan-Simon witness, the state can be arbitrarily close to a separable one and $p(x_1,x_2)$ essentially correspond to classical correlations \footnote{To illustrate that, consider a state $\rho_\epsilon$ (\ref{state in x}) with  $f_1(x, \bar x) = f_2(x, \bar x) =1$ for $|x-\bar x| <\epsilon$  and zero otherwise. The peculiarity of this state is that $f$ satisfies \unexpanded{$\mn{f'} = \mn{f''} = 0$} in such a way that the violation of the Duan-Simon criterion by $\rho_\epsilon$ is independent of $\epsilon$ unless it is strictly equal to zero. Therefore $\rho_\epsilon$ with $\epsilon$ approaching zero is an exemple of a state that can give an arbitrarily high violation of the Duan criteria, while being arbitrarily close to a separable state.}. However, under the assumption of a Gaussian decay of coherence $f(x_1,\bar x_1,x_2,\bar x_2) = e^{-(x_1-\bar x_1)^2 / (2 \gamma_1^2)} e^{-(x_2-\bar x_2)^2 / (2 \gamma_2^2)},$ Eq. (\ref{bound on f}) becomes
$
\frac{1}{\gamma_1^2} + \frac{1}{\gamma_2^2} \leq V(\bar X^{\frac{\pi}{2}}_1 + \bar X^{\frac{\pi}{2}}_2).
$
This implies $\text{min}(\gamma_1, \gamma_2) \geq 1/ \sqrt{V(\bar X^{\frac{\pi}{2}}_1 + \bar X^{\frac{\pi}{2}}_2)}$, i.e., if one observes the variance $V_{\psi_{\text{tms}}}(\bar X^{\frac{\pi}{2}}_1 + \bar X^{\frac{\pi}{2}}_2)$ of the total momentum, we can certify that the correlations $|p(x_1,x_2)|^2$ are quantum at least in the range 
\eq
\label{eq:8}
x_C =\frac{1}{\sqrt{V_{\psi_{\text{tms}}}(\bar X^{\frac{\pi}{2}}_1 + \bar X^{\frac{\pi}{2}}_2)}}.
\eeq
Accordingly if the coherence range $x_C$ is lower than the correlation range as witnessed by $\sigma_{\max}$, one can only claim that the state exhibits quantum correlations within the range $x_C$, which is then the true size of the state. Revealing the size of large quantum states thus requires to reveal narrow variances which is harder and harder as the size increases, cf. below.\\

\paragraph{General measures for multimode cv states ---}
Besides measures for macroscopic distinguishability, there has been recent proposals that aim to go beyond the basic structure $| \phi_0 \rangle + \left| \phi_1 \right\rangle $ \cite{Shimizu_stability_2002, Shimizu_Detection_2005, Lee_Quantification_2011, Frowis_Measures_2012}. While the measures of Refs. \cite{Shimizu_stability_2002, Shimizu_Detection_2005, Frowis_Measures_2012} were originally defined for spin systems, the definition of Ref.~\cite{Lee_Quantification_2011} is directly suitable for cv photonic states. For pure states, these three proposals are comparable since a state $| \psi \rangle $ is called macroscopically quantum if it shows a large variance $V$ with respect to a restricted class of operators. In the spin case, the proposals \cite{Shimizu_stability_2002, Shimizu_Detection_2005, Frowis_Measures_2012} focus on sums of local operators (henceforth simply called ``local operators''), whereas Lee and Jeong \cite{Lee_Quantification_2011} define their measure for pure states proportional to $V(\bar{X}^0) + V(\bar{X}^{\pi/2})$. In Ref.~\cite{Frowis_Linking_2015}, it was argued that local operators in the spin case play to some extent the same role as quadature operators in mono-mode photonic systems.

The common feature of the proposals for mixed states is that the measures \cite{Shimizu_stability_2002, Shimizu_Detection_2005, Lee_Quantification_2011, Frowis_Measures_2012} are convex in the state, which is an important and natural feature for the present purpose. There are no clear arguments in favor of one of the proposal. Nevertheless, we focus here on the quantum Fisher information (QFI) \cite{Braunstein_Statistical_1994}; denoted as $\mathcal{F}_{\rho}(\bar{X})$ for the state $\rho$ and the operator $\bar{X}$. Importantly, the QFI is the convex roof of the variance \cite{Toth_Extremal_2013,Yu_Quantum_2013} (up to a factor four), that is, it is the largest convex function that reduces to the variance for pure states. For experiments, it is interesting to note that there exist lower bounds on the QFI based on measurable quantities \cite{Frowis_Tighter_2014}.

The extension to photonic states with $n>1$ modes is not straightforward. Indeed, a multimode version for the measure of Lee and Jeong was proposed \cite{Lee_Quantification_2011}. However, it is additive and hence a bunch of ``kitten states'' $\left| \psi_{\alpha} \right\rangle ^{\otimes n} \propto( \left| \alpha \right\rangle + \left| -\alpha \right\rangle )^{\otimes n}$ (with potentially small $\alpha$ but large $n$) is as macroscopically quantum as a ``big'' single cat state $\left| \psi_{\sqrt{n}\alpha} \right\rangle \propto | \sqrt{n} \alpha \rangle + \left| -\sqrt{n} \alpha \right\rangle $. Here, we propose instead to use a similar account that has been successfully applied in the spin case \cite{Shimizu_stability_2002, Shimizu_Detection_2005, Frowis_Measures_2012}. The idea is that the effective size of a product state is the average value of its components, while entangled states should be able to profit from quantum correlations between the modes. Both requirements are achieved by defining the effective size for $\rho$ as
\begin{equation}
\label{eq:1}
N_{\mathrm{eff}}(\rho) = \frac{1}{4n} \max_{\boldsymbol{\theta}} \mathcal{F}_{\rho}(X_{\boldsymbol{\theta}}),
\end{equation}
where $X_{\boldsymbol{\theta}} = \sum_{i=1}^n \bar{X}_i^{\theta_i}$. In words, one maximizes the QFI (or the variance for pure states) with respect to sums of local quadrature operators parametrized by  $\boldsymbol{\theta} = (\theta_1,\dots,\theta_n)$. The examples from above then lead to $N_{\mathrm{eff}}(\left| \psi_{\alpha} \right\rangle ^{\otimes n}) = 2 |\alpha|^2 /[1 + \exp(-2 |\alpha|^2)]$ and $N_{\mathrm{eff}}(\left| \psi_{\sqrt{n}\alpha} \right\rangle) = 2n|\alpha|^{2}/[1 + \exp(-2n |\alpha|^2)]$ (cp.~to \cite{Volkoff_Measurement-_2014}).

We now come to the evaluation of the effective size for the two-mode squeezed vacuum state. It is simple to see that the variance is largest for the quadratures that are maximally correlated. For the state (\ref{2modesqueezed}), these are the operators $\bar{X}^0_1 + \bar{X}^0_2$ and $\bar{X}^{\pi/2}_1 + \bar{X}^{\pi/2}_2$. The effective size for each of these choices reads $N_{\mathrm{eff}}(\psi_{\mathrm{tms}}) = \frac{1}{2} V(\bar{X}^0_1 + \bar{X}^0_2) = \frac{1}{2}e^{2g} \approx N$ which is approximately half of the value as for the cat state with the same photon number, $N_{\mathrm{eff}}(\left| \psi_{\alpha} \right\rangle ) \approx 2 N$.
 
In principle, the effective size of a pure state could be determined by witnessing a large variance for sums of quadrature operators. However, for mixed states, a large variance is not sufficient. Instead, one has to verify a large value of a convex function like the QFI. Since this quantity is typically only accessibly through a full state tomography, one has to find other means to estimate it. Recently, a general lower bound on the QFI has been found \cite{Frowis_Tighter_2014}. It was shown that for any quantum state $\rho$ and any pair of operators $A,B$, it holds that $V_{\rho}(A) \mathcal{F}_{\rho}(B) \geq \langle i [A,B] \rangle_{\rho}^2$, which is a tighter version of the Heisenberg uncertainty relation. Here, we use this inequality to bound the QFI from below. For $B = \bar{X}^0_1 + \bar{X}^0_2$, we set $A = \bar{X}^{\pi/2}_1 + \bar{X}^{\pi/2}_2$ and find $i[A,B] = -2$. Hence one has 
\begin{equation}
\label{eq:2}
N_{\mathrm{eff}}(\rho) \geq \frac{1}{V_{\rho}(\bar{X}^{\pi/2}_1 + \bar{X}^{\pi/2}_2)}.
\end{equation} 
For the two-mode squeezed state, the anti-correlations between $\bar{X}^{\pi/2}_1 $ and $ \bar{X}^{\pi/2}_2$ lead to a reduced variance and therefore to a potentially large value of $N_{\mathrm{eff}}$.

Note that Eq.~(\ref{eq:2}) [as well as Eq.~(\ref{eq:8})] resembles very much the ideas of Refs.~\cite{Cavalcanti_Signatures_2006,Cavalcanti_Criteria_2008} for a generalized notion of macroscopic quantum coherences. However, we use these expressions only to bound quantitative measures. These measures are general enough to compare two-mode squeezed states with other states like cat states.
\\

\paragraph{On the difficulty to certify the quantum nature of two-mode squeezed states ---} The common feature of measures for macroscopicity presented before is the requirement to reveal narrow variances, especially when dealing with large size states. How hard is it in practice? To answer this question, we consider the effect of various experimental imperfections on the observed variance $V_{\psi_{\mathrm{tms}}}(\bar{X}^{\pi/2}_1 + \bar{X}^{\pi/2}_2)$.
 
(i) Consider first a noise along $\bar{X}^{0}$ which acts on a state $\rho$ as  
$\rho \mapsto \int d\lambda \, h(\lambda) e^{i \hat X_0 \lambda} \rho e^{- i \hat X_0 \lambda} $ with characteristic function (noise distribution) $h(\lambda)$ of variance $\Delta^2 h$. The effect of this noise can be directly absorbed in the statistics of the momentum distribution and leads to the following modification of the variance
$
V(\bar{X}^{\pi/2}_1 + \bar{X}^{\pi/2}_2) \mapsto V(\bar{X}^{\pi/2}_1 + \bar{X}^{\pi/2}_2) + \Delta^2 h_1 +\Delta^2 h_2.
$
Therefore, if the experimental setup suffers from such a noise, we cannot certify state of effective size larger than 
$
N_\text{eff}^{\max}= \frac{1}{ \Delta^2 h_1 + \Delta^2 h_2}.
$ 

(ii) Similarly, consider a loss channel with transmission $\eta.$ It leads to 
$V(\bar{X}^{\pi/2}_1 + \bar{X}^{\pi/2}_2)  \mapsto \eta V(\bar{X}^{\pi/2}_1 + \bar{X}^{\pi/2}_2)+  (1-\eta)$ and
the maximal certifiable size is given by 
$
N_\text{eff}^{\max}= \frac{1}{1- \eta}.
$

(iii) Now consider a phase noise characterized by the variance $\Delta \varphi^2 =\int p(\varphi) \varphi^2 d\varphi.$ It increases the observed variance according to 
$V(\bar{X}^{\pi/2}_1 + \bar{X}^{\pi/2}_2)  \geq \Delta \varphi^2  (\mn{\bar{X}^{0}_1} + \mn{\bar{X}^{0}_2} ) $. Specifically for the two-mode squeezed state, one has
$
{N_\text{eff}^{\max}}({\psi_{\mathrm{tms}}})= \frac{1}{ \Delta \varphi^2  (2 \sinh^2(g)+1)}
$
which decays exponentially with the squeezing parameter (in the limit of large enough $g$).

\begin{figure}
\includegraphics[width=0.4\textwidth]{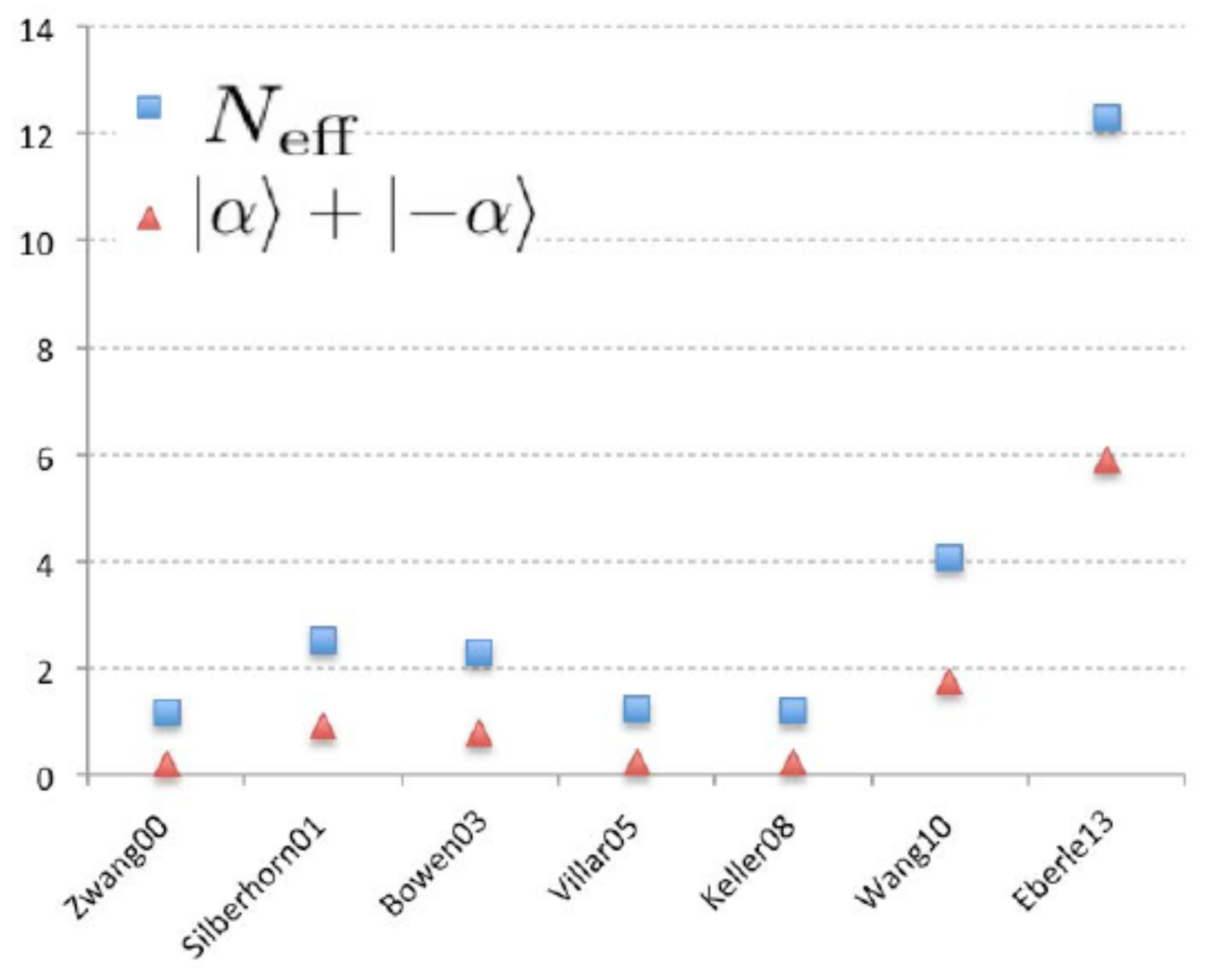}
\caption{Bounds on the effective size $N_\text{eff}$ (blue squares) of two-mode squeezed states obtained from experimental data reported in Refs.~\cite{Zhang_Experimental_2000, Silberhorn_Generation_2001, Bowen_Experimental_2003, Villar_Generation_2005, Keller_Experimental_2008, Wang_Experimental_2010, Eberle_Experimental_2013} using inequality (\ref{eq:2}). The red triangles indicate the minimal photon number $N$ necessary for a cat state $\ket{\alpha}+\ket{-\alpha}$ to have the same effective size according to Eq.~(\ref{eq:1}). For example, the state reported in Ref.~\cite{Zhang_Experimental_2000} has a size $N_{\text{eff}} \geq 1.2$ for which one needs at least a cat state with $N \approx 0.2$ for the same size.}
\label{Fig1}
\end{figure} 
 
In each case, we clearly see that it becomes harder and harder to observe narrow variances with two-mode squeezed states as their size increases. This is in agreement with recent results \cite{Frowis_Certifiability_2013, Sekatski_How_2014} stating that it is difficult to observe the quantum nature of macroscopic states. This naturally raises the question of the size of states that can be observed in practice. Note first that formulas (\ref{eq:8}) and (\ref{eq:2}) are general, i.e.~the variance along the conjugate of quadratures that are maximally correlated gives a bound on the size of the measured state. We use them to compare the size of states obtained in various setups in which the $\chi^2$ nonlinearity is either seeded or embedded in a cavity \cite{Zhang_Experimental_2000, Silberhorn_Generation_2001, Bowen_Experimental_2003, Villar_Generation_2005, Keller_Experimental_2008, Wang_Experimental_2010, }, see Fig.~\ref{Fig1}. 
All these experiments have in common that the Duan-Simon criterion is used to reveal entanglement and the photon number is large. We clearly see that their size cannot be compared to their mean photon number. In the seeded case, the reason is that the seed increases the photon number but does not change the variance. Similarly with a cavity, the photon number can be large even if the gain slightly dominates the loss provided that the cavity finesse is large but the variance of interest is limited by the ratio between the gain and the loss only (see Appendix B). Interestingly, the results presented in Fig. \ref{Fig1} can be used directly to quantitatively estimate the metrologic usefulness of states realized experimentally as the size $N_{\text{eff}}$ gives the QFI through the formula (\ref{eq:1}).\\

\paragraph{Acknowledgements ---} We thank M. Mitchell for sharing several discussions with us, one of them having initiated this work. We also thank warmly W. D\"ur, J. Laurat and M. Skotiniotis for many discussions. This work was supported by the National Swiss Science Foundation (SNSF), Grant number PP00P2-150579 and  P2GEP2-151964, the Austrian Science Fund (FWF), Grant number J3462 and P24273-N16 and the European Research Council (ERC MEC).\\

\paragraph{Appendix A: Coherence length of correlation}

Every two-mode state $\rho$ can be expressed in the joint $x$-basis
\eq 
\rho = \int F(x_1, x_2, \bar x_1, \bar x_2) \coh{x_1, x_2}{\bar x_1, \bar x_2} d\vec x.
\eeq
However for our purpose it is useful to make the decomposition
\eq
 F(x_1, x_2, \bar x_1, \bar x_2) = p(x_1,x_2) p^*(\bar x_1,\bar x_2) f(x_1,x_2, \bar x_1, \bar x_2),\label{decomposition}
\eeq 
where we can enforce that $\int |p(x_1,x_2)|^2 d x_1 d x_2 =1$ and  $f(x_1= \bar x_1,x_2 = \bar x_2) = 1$,  and consequently $|f(x_1\neq \bar x_1,  x_2 \neq \bar x_2)| \leq 1$.
The later inequality is ensured by positivity of $\rho$, i.e. if it does not hold then there is a state $\alpha \ket{x_1,x_2} + \beta \ket{\bar x_1, \bar x_2}$ that 
has a negative overlap with $\rho$. The decomposition \eqref{decomposition} is useful, because the function $f(x_1, x_2, \bar x_1, \bar x_2)$ can be simply interpreted as characterizing the lack of purity of $\rho$, since for 
a pure state $\ket{\psi}= \int p(x_1, x_2) \ket{x_1,x_2} dx_1 d x_2$ it satisfies $ f(x_1, x_2, \bar x_1, \bar x_2)\equiv 1$.

Let us now consider the mean values of $\mn{p_{1(2)}}$, $\mn{p_{1(2)}^2}$ and $\mn{p_1 p_2}$ on the state $\rho$. In this section we denote $\bar X^0=x$ and $\bar X^{\pi/2}= p$ Using the representation of momenta eigenstate   in the $x$ basis $\ket{p}=\frac{1}{\sqrt{2\pi}}\int e^{i x p} \ket{x}$  one gets
\begin{align}
&\mn{p_1} = \int F(x_1, x_2, \bar x_1, \bar x_2) p  \frac{e^{- i p(x_1-\bar x_1)}}{2\pi} \delta(x_2-\bar x_2) dp d\vec x \nonumber \\ \nonumber
&\mn{p_1^2} = \int F(x_1, x_2, \bar x_1, \bar x_2) p^2  \frac{e^{- i p(x_1-\bar x_1)}}{2\pi} \delta(x_2-\bar x_2) dp d\vec x\\ \nonumber
&\mn{p_1 p_2} = \int F(x_1, x_2, \bar x_1, \bar x_2) p_1 p_2  \frac{e^{- i p_1(x_1-\bar x_1)}}{2\pi}\times\\ &\frac{e^{- i p_2(x_2-\bar x_2)}}{2\pi} dp_1 dp_2 d\vec x.
\end{align}
Using $p^n e^{- i p( \Delta x)} =  (i \partial_{\Delta x})^n e^{- i p( \Delta x)}$ and $\frac{1}{2\pi} \int e^{- i p (x- \bar x)} dp = \delta(x-\bar x)$ a simple integration by parts allows to rewrite the above expressions as
\begin{align}
&\mn{p_1} = \int    (i \partial_{x_1 - \bar x_1}) F(x_1, x_2, \bar x_1, \bar x_2)|_{x_1= \bar x_1} dx_1 dx_2 \nonumber \\ \nonumber
&\mn{p_1^2} =  \int    (i \partial_{x_1 - \bar x_1})^2 F(x_1, x_2, \bar x_1, \bar x_2)|_{x_1= \bar x_1} dx_1 dx_2\\ \nonumber
&\mn{p_1 p_2} = \int   (i \partial_{x_2 - \bar x_2}) (i \partial_{x_1 - \bar x_1})\\ \nonumber& F(x_1, x_2, \bar x_1, \bar x_2)|_{x_1= \bar x_1 , x_2 = \bar x_2} dx_1 dx_2.
\end{align} 
Those expressions allow to use the decomposition \eqref{decomposition} to its full advantage, leading to
\begin{align}\label{means}
&\mn{p_1} = \mn{p_1}_{f\equiv 1} \\ \nonumber
&\mn{p_1^2} =  \mn{p_1^2}_{f\equiv 1} - \mn{\partial_{x_1-\bar x_1}^2 f}\\ \nonumber
&\mn{p_1^2} =  \mn{p_2^2}_{f\equiv 1} - \mn{\partial_{x_2-\bar x_2}^2 f}\\ \nonumber
&\mn{p_1 p_2} = \mn{p_1 p_2}_{f\equiv 1} - \mn{\partial_{x_1-\bar x_1}\partial_{x_2-\bar x_2} f},
\end{align} 
with the averages $\mn{ }_{f\equiv1}$ are taken over the pure state $\ket{\psi} =\int p(x_1, x_2) \ket{x_1,x_2}dx_1 dx_2$  and 
\eq
\mn{D[f]}=\int |p(x_1,x_2)|^2 D[f](x_1,x_2) dx_1 dx_2.
\eeq
Notice that to derive these expressions, we used the fact that the first derivatives of $f$ are zero (because $\rho$ is hermitian).
 
The expressions \eqref{means} allow one to rewrite the variance of $p_1+p_2$ in a form where the contributions of $p(x_1,x_2)$ and $f(x_1,x_2,\bar x_1, \bar x_2)$ are separated
\eq
V(p_1+p_2) = V(p_1+p_2)_{f\equiv 1} - \mn{ (\partial_{x_1-\bar x_1}+\partial_{x_2-\bar x_2})^2 f}.
\eeq
Remark that for pure states, the variance $V(p_1+p_2)_{f\equiv 1}$ is always positive. This allows us to upper bound the decay of coherences in the $x$-basis
\eq\label{bound}
 - \mn{ (\partial_{x_1-\bar x_1}+\partial_{x_2-\bar x_2})^2 f} \leq V(p_1+p_2).
\eeq
Without supplementary assumptions local derivatives of $f$ at $x_1 = \bar x_1$ and $x_2 = \bar x_2$ are not sufficient to determine global properties, such that the variances $V(x_1 - \bar x_1)$ and $V(x_2 - \bar x_2)$  of f, as one can imagine irregular functions $f$ that have zero derivatives but arbitrarily small variance   $V(x_1 - \bar x_1)$ (e.g. the step function of arbitrarily small width). But assuming a Gaussian profile for the decay of coherence allows us to draw conclusion on the coherence width of $f$ from the upper bound \eqref{bound}, as we show in the main text.

\paragraph{Appendix B : Quadrature correlations for an amplifier with loss}

In this section we derive a simple model for a two mode optical parametric amplification in a cavity with loss. The amplification Hamiltonian is given by
\eq
H_A= i \chi (a^\dag b^\dag - a b),
\eeq
with $\chi >0$. The loss is described by a beam-splitter operating on each mode $a$ and $b.$ The global process can be seen as a sequence of alternating infinitesimal amplifiers with gain $\chi dt$ and losses with intensity transmission $1- 2 \lambda dt$. Consider an operator of the form
\eq\label{O operator}
\mathcal{O}(\eta, \mu, \kappa) = e^{\kappa} e^{i (\eta a^\dag + \mu b^\dag)} e^{i (\eta^* a + \mu^* b)}
\eeq
and propagate it through an elementary step of our process (amplification + loss). It is easy to see that after a infinitesimal time step $dt$ the operator becomes
\begin{align}
tr_{loss} U_{dt}^\dag \mathcal{O}(\eta, \mu, \kappa) U_{dt} = \\ \nonumber
 \mathcal{O}
(\eta +(\mu^* \chi - \eta \lambda )dt ,\nonumber\\ \mu +(\eta^* \chi - \mu \lambda  ) dt, \kappa - ( \mu^* \eta^*  +\mu \eta)\chi dt),
\end{align}
where we omit terms of higher order in $dt$ in the exponent and the trace is there to remind that the loss is not a unitary evolution (the trace is taken over the vacuum modes of the environment). So during the evolution the operator $\mathcal{O}_t = \mathcal{O}(\eta(t), \mu(t), \kappa(t))$ keeps its form while the scalar functions satisfy the system of differential equations
\eq
\begin{cases}
\dot \eta(t) =\chi \mu^*(t) - \lambda \eta(t) \\
\dot \mu(t) =\chi \eta^*(t) - \lambda \mu(t) \\
\kappa(t) = -\chi \int_0^t \big(\eta^*(s) \mu^*(s)+\eta(s)\mu(s) \big)ds + \kappa(0).
\end{cases}
\eeq
The solution is straightforward
\begin{align}\label{time solution}&
\left(\begin{array}{c}
\eta(t)\\
\mu^*(t)
\end{array}\right) =
\exp \big(\left(
\begin{array}{cc}
-\lambda & \chi \\
\chi & - \lambda
\end{array}
\right ) t\big)
\left(\begin{array}{c}
\eta_0\\
\mu^*_0
\end{array}\right)\\& \nonumber
\left(\begin{array}{c}
\mu(t)\\
\eta^*(t)
\end{array}\right) =
\exp \big(\left(
\begin{array}{cc}
-\lambda & \chi \\
\chi & - \lambda
\end{array}
\right ) t\big)
\left(\begin{array}{c}
\mu_0\\
\eta^*_0
\end{array}\right)
\\&
\kappa(t) =
\left(\begin{array}{cc}
\eta_0 & \mu^*_0
\end{array}\right)\cdot
\Big( \int_0^t
e^{ \left(
\begin{smallmatrix}
-\lambda & \chi \\
\chi & - \lambda
\end{smallmatrix}
\right ) 2s }
ds\Big)\cdot
\left(\begin{array}{c}
\mu_0\\
\eta^*_0 
\end{array}\right)  +\kappa_0 \nonumber.
\end{align}

Given the expression of the propagated $\mathcal{O}_t$ operator one can evaluate the quadrature statistics of an evolved state. Let us calculate the following probability  $p(x_\theta^a, y_\xi^b)= \mn{ \prjct{x_\theta^a, y_\xi^b} }$ on an evolved state. Using
\begin{align}
\prjct{x_\theta^a} = \frac{1}{2\pi} \int d\zeta e^{i \zeta (\hat x_\theta^a -x)} d\zeta\\
\hat x_\theta^a = \frac{1}{\sqrt{2}}(a \, e^{-i \theta} + a^\dag e^{ i \theta}),
\end{align}
the projector on quadrature eigenstates can be expressed as
\begin{align}
\prjct{x_\theta^a, y_\xi^b} =\int \frac{d\zeta d\gamma}{(2\pi)^2} e^{- i \zeta x - i \gamma y }  \times \nonumber\\
\underbrace{e^{-\frac{\zeta^2 + \gamma^2}{4}}
e^{i(\frac{\zeta e^{i\theta}}{\sqrt{2}}a^\dag +\frac{\gamma e^{i\xi}}{\sqrt{2}}b^\dag) }
e^{i(\frac{\zeta e^{-i\theta}}{\sqrt{2}}a +\frac{\gamma e^{-i\xi}}{\sqrt{2}}b) }}_{\mathcal{O}_0}
\end{align}
where the non-trivial part has the form of the operator $\mathcal{O}$ in \eqref{O operator} with $\eta_0 = \frac{\zeta e^{i\gamma}}{\sqrt{2}}$,
$\mu_0 = \frac{\gamma e^{i\xi}}{\sqrt{2}}$ and $\kappa_0 =- \frac{\zeta^2+ \gamma^2}{4}$. Resolving the time evolution of $\mathcal{O}_t$ using \eqref{time solution}, one gets that for the final probability (at time t)
\eq
p(x_\theta^a, y_\xi^b) = \int \frac{d\zeta d\gamma}{(2\pi)^2} e^{- i \zeta x - i \gamma y }  \mn{\mathcal{O}_t}.
\eeq
For a coherent input states (seeds) the mean value $\mn{\mathcal{O}_t} =\bra{\alpha, \beta} \mathcal{O}_t \ket{\alpha,\beta} = e^{\kappa(t)}e^{i(\alpha^* \eta(t) + \beta^*\mu(t))} e^{i(\alpha \eta^*(t) + \beta\mu^*(t))} $ is particularly simple. 

 A direct calculation using the formulas above gives
\begin{widetext}
\begin{align}
p(x_\theta^a, y_\xi^b) =\int \frac{d\zeta d\gamma}{(2\pi)^2} 
e^{- i \left(\begin{smallmatrix}
\zeta & \gamma
\end{smallmatrix}\right)
\cdot
\left(\begin{smallmatrix}
x - Z(\alpha, \beta)\\ y - \Gamma(\alpha, \gamma)
\end{smallmatrix}\right)} e^{-\frac{\zeta^2 + \gamma^2}{4}}\times\\ \nonumber 
e^{-\frac{1}{2}
\left(\begin{smallmatrix}
\zeta & \gamma
\end{smallmatrix}\right)\cdot
\left(
\begin{smallmatrix}
 \frac{1}{4} \left(\frac{-1+e^{-2 t (\lambda +\chi )}}{\lambda +\chi }-\frac{-1+e^{2 t (\chi -\lambda )}}{\lambda -\chi }\right) & e^{i (\theta +\xi )} \left(-\frac{-1+e^{2 t (\chi -\lambda )}}{4 (\lambda -\chi )}-\frac{-1+e^{-2 t (\lambda +\chi )}}{4 (\lambda +\chi )}\right) \\
 e^{-i (\theta +\xi )} \left(-\frac{-1+e^{2 t (\chi -\lambda )}}{4 (\lambda -\chi )}-\frac{-1+e^{-2 t (\lambda +\chi )}}{4 (\lambda +\chi )}\right) & \frac{1}{4} \left(\frac{-1+e^{-2 t (\lambda +\chi )}}{\lambda +\chi }-\frac{-1+e^{2 t (\chi -\lambda )}}{\lambda -\chi }\right) \\
\end{smallmatrix}
\right)\cdot
\left(\begin{smallmatrix}
\zeta \\ \gamma
\end{smallmatrix}\right)}
\end{align}
\end{widetext}
The Fourier transform yields a Gaussian joint probability
\eq
p(x_\theta^a, y_\xi^b)= \frac{\sqrt{r_- r_+}}{4 \pi }
e^{-\frac{1}{4}
\left(\begin{smallmatrix}
x-Z(\alpha,\beta) \,& y-\Gamma(\alpha, \beta)
\end{smallmatrix}\right)\cdot
M\cdot
\left(\begin{smallmatrix}
x -Z(\alpha, \beta)\\ y-\Gamma(\alpha, \beta)
\end{smallmatrix}\right)},
\eeq
where $r_+$ and $r_-$ are the two eigenvalues of the matrix $M$ given by
\begin{align}
r_- = \left(\frac{\lambda +\chi  e^{-2 t (\lambda +\chi )}}{4 (\lambda +\chi )}\right)^{-1}\\
r_+= \left( \frac{\lambda -\chi  e^{2 t (\chi -\lambda )}}{4 (\lambda -\chi )}\right)^{-1}.
\end{align} 
Accordingly the joint probability $p(x_\theta^a, y_\xi^b)$ decomposes in a product of two Gauissians with variance $\Delta_- = 2/r_-$ in the squeezed direction (decreasing with time) and $\Delta_+ =2/r_+$ in the anti-squeezed direction (increasing with time). Let us comment on their asymptotic values for $t \to\infty$ (limit of high finesse) for the two different regimes:  
\begin{itemize}
\item \emph{Below threshold} $\lambda>\chi$,
\eq
\Delta_+\to \frac{\lambda}{2(\lambda- \chi)} \quad  \Delta_-\to \frac{\lambda}{2(\lambda+ \chi)}
\eeq
both variances saturate at constant values.
\item \emph{Above threshold} $\lambda<\chi$,
\eq
\Delta_+\to \frac{\chi}{2(\chi-\lambda)}e^{2t(\chi-\lambda)} \quad  \Delta_-\to \frac{\lambda}{2(\lambda+ \chi)}
\eeq
while the variance in the anti-squeezed direction increases exponentially with time (finesse), the squeezed width can not be decreased below a constant $\frac{1}{2} \frac{1}{1+ \chi/\lambda}$ set by the quality factor of the amplification process $\frac{\chi}{\lambda}$.
\end{itemize}

\bibliographystyle{apsrev4-1}
\bibliography{TwoModeSqueezedStates.bib}

\end{document}